\documentclass[aps,physrev,twocolumn,superscriptaddress]{revtex4-2}

\usepackage{graphicx}%
\usepackage{multirow}%
\usepackage{amsmath,amssymb,amsfonts}%
\usepackage{amsthm}%
\usepackage{mathrsfs}%
\usepackage{xcolor}%
\usepackage{textcomp}%
\usepackage{booktabs}%
\usepackage{listings}%
\usepackage{todonotes}
\usepackage{lineno}
\usepackage{xspace}

\begin{document}

\newcommand{\trackml}{\texttt{TrackML}\xspace}
\newcommand{\sphenix}{\texttt{sPHENIX}\xspace}

\newcommand{\geant}{\textsc{Geant4}\xspace}
\newcommand{\pythia}{\textsc{Pythia8}\xspace}

\newcommand{\checksymbol}{\textcolor{green}{\ding{51}}} 
\newcommand{\crosssymbol}{\textcolor{red}{\ding{55}}}   

\newcommand{\maskformer}{\texttt{MaskFormer}\xspace}
\newcommand{\masktwoformer}{\texttt{Mask2Former}\xspace}
\newcommand{\eggnet}{\texttt{EggNet}\xspace}
\newcommand{\exatrkx}{\texttt{Exa.TrkX}\xspace}
\newcommand{\fm}{\texttt{FM4NPP}\xspace}
\newcommand{\fmhead}{\texttt{Adapter Only}\xspace}
\newcommand{\GATConv}{\texttt{GATConv}\xspace}
\newcommand{\GCNConv}{\texttt{GCNConv}\xspace}
\newcommand{\GraphConv}{\texttt{GraphConv}\xspace}
\newcommand{\SAGEConv}{\texttt{SAGEConv}\xspace}
\newcommand{\oneformer}{\texttt{OneFormer3D}\xspace}

\newcommand{\dedx}{\ensuremath{dE/dx}\xspace}

\newcommand{\gev}[1]{\SI{#1}{\giga\electronvolt}}
\newcommand{\mev}[1]{\SI{#1}{\mega\electronvolt}}
\newcommand{\pt}{\ensuremath{p_{\mathrm{T}}\xspace}}
\newcommand{\GEANT}{\textsc{Geant4}\xspace}

\newcommand{\pointembeds}{spacepoint embeddings\xspace}
\newcommand{\tqueries}{track queries\xspace}
\newcommand{\tquery}{track query\xspace}
\newcommand{\rtqueries}{refined track queries\xspace}
\newcommand{\tembed}{track embedding\xspace}
\newcommand{\tembeds}{track embeddings\xspace}
\newcommand{\tinspreds}{track instance predictions\xspace}
\newcommand{\tinspred}{track instance prediction\xspace}
\newcommand{\td}{transformer decoder\xspace}

\newcommand{\paren}[1]{\left(#1\right)}
\newcommand{\set}[1]{\left\{#1\right\}}
\newcommand{\bracket}[1]{\left[#1\right]}
\newcommand{\abs}[1]{\left|#1\right|}
\newcommand{\norm}[1]{\left\Vert#1\right\Vert}

\newcommand{\addcitation}{\textcolor{red}{[ADD CITATION]}\xspace}
\newcommand{\pleasecheck}{\textcolor{red}{[PLEASE CHECK]}\xspace}
\newcommand{\adddetails}[1]{\textcolor{green}{ADD DETAILS on {#1}}\xspace}

\definecolor{BrickRed}{rgb}{0.8, 0.25, 0.33}
\definecolor{DarkGreen}{rgb}{0.0, 0.4, 0.0}
\definecolor{RoyalBlue}{rgb}{0.25, 0.41, 0.88}
\definecolor{Orchid}{rgb}{0.85, 0.44, 0.84}
\definecolor{ProcessCyan}{rgb}{0.0, 0.68, 0.93}
\definecolor{DarkOrange}{rgb}{1.0, 0.55, 0.0}
\definecolor{Brown}{rgb}{0.65, 0.16, 0.16}
\definecolor{OliveGreen}{rgb}{0.33, 0.42, 0.18}
\definecolor{BlueViolet}{rgb}{0.54, 0.17, 0.89}
\definecolor{MidnightBlue}{rgb}{0.1, 0.1, 0.44}
\definecolor{Goldenrod}{rgb}{0.85, 0.65, 0.13}
\newcommand{\ray}[1]{\textcolor{BrickRed}{[Ray: #1]}}
\newcommand{\david}[1]{\textcolor{DarkGreen}{[David: #1]}}
\newcommand{\jin}[1]{\textcolor{RoyalBlue}{[Jin: #1]}}
\newcommand{\shuhang}[1]{\textcolor{Orchid}{[Shuhang: #1]}}
\newcommand{\shinjae}[1]{\textcolor{ProcessCyan}{[Shinjae: #1]}}
\newcommand{\yi}[1]{\textcolor{DarkOrange}{[Yi: #1]}}
\newcommand{\joe}[1]{\textcolor{Brown}{[Joe: #1]}}
\newcommand{\haiwang}[1]{\textcolor{OliveGreen}{[Haiwang: #1]}}
\newcommand{\yeonju}[1]{\textcolor{BlueViolet}{[Yeonju: #1]}}
\newcommand{\xihaier}[1]{\textcolor{MidnightBlue}{[Xihaier: #1]}}
\newcommand{\yuewei}[1]{\textcolor{Goldenrod}{[Yuewei: #1]}}

\title{Track Finding and Particle Identification Performance with Particle Physics Foundation Models}


\author{Shuhang Li}
\affiliation{Columbia University, New York, NY 10027, USA}
\author{Haiwang Yu}
\author{Gabor Galgoczi}
\affiliation{Brookhaven National Laboratory, Upton, NY 11973, USA}
\author{Yeonju Go}
\affiliation{Brookhaven National Laboratory, Upton, NY 11973, USA}
\affiliation{State University of New York, Stony Brook, New York, NY 11794, USA}
\author{Yi Huang}
\affiliation{Brookhaven National Laboratory, Upton, NY 11973, USA}
\author{David Park}
\affiliation{Brookhaven National Laboratory, Upton, NY 11973, USA}
\author{Michael Peters}
\affiliation{Massachusetts Institute of Technology, Cambridge, MA 02139, USA}
\author{Christopher Pinkenburg}
\affiliation{Brookhaven National Laboratory, Upton, NY 11973, USA}
\author{Christof Roland}
\affiliation{Massachusetts Institute of Technology, Cambridge, MA 02139, USA}

\author{Jin Huang}
\author{Yihui Ren}
\author{Joseph D. Osborn}
\affiliation{Brookhaven National Laboratory, Upton, NY 11973, USA}



\date{\today}

\begin{abstract}
Modern high energy nuclear and particle physics experiments produce large amounts of sparse data that must be reduced significantly in size before performing physics analysis. Recently, there has been an effort to evaluate Foundation Models as a potential tool for performing data analysis. These models have been shown to learn features in high dimensional datasets based on the multi-modal data they are trained on. To show their effectiveness at performing complex pattern recognition tasks, such as reconstructing charged particles in a magnetic field, we evaluated the performance of a Foundation Model based reconstruction on several downstream tasks with a dataset composed of spacepoints in a Time Projection Chamber. We show that the performance of the Foundation Model based data analysis significantly improves upon that of traditional methods, and furthermore can be adapted to learn additional tasks that are not necessarily related to the initial training objective.
\end{abstract}

\maketitle


\section{Introduction}
Data reconstruction for high energy nuclear and particle physics experiments is becoming increasingly challenging as modern and future facilities deliver ever higher luminosities. At the Relativistic Heavy Ion Collider, the sPHENIX experiment operates with a streaming readout data acquisition at rates far beyond those of previous generations of RHIC detectors, while the High-Luminosity Large Hadron Collider and the future Electron-Ion Collider will present comparable or greater demands, which increase the detector occupancies. Consequently, there has been significant effort in exploring new technologies to maintain high physics performance.   

In artificial intelligence (AI) research, the emergence of large language models has marked a major shift in model development, moving from specialized architectures designed for a single task toward general architectures adaptable to many tasks~\cite{Vaswani:2017}. Foundation models (FMs) are pretrained on large datasets to learn general representations, and are subsequently adapted to a wide variety of downstream tasks through lightweight fine-tuning or small task-specific adapters.  
Their ability to use self-supervised learning is particularly attractive for nuclear and particle physics, where unlabeled experimental data are abundant while fully labeled simulated data are computationally expensive and carry their own modeling uncertainties.

There has been a growing research effort into utilizing FMs for nuclear and particle physics, with much of the development targeting jet physics, since jets can be represented as ordered sets of constituents that resemble the token sequences of language models. Models pretrained on large jet-constituent datasets have been adapted to classification, generation, regression, reweighting, unfolding, and anomaly detection within a single framework~\cite{Birk:2024OmniJet,Golling:2024MPM,Harris:2025RS3L,Mikuni:2025OmniLearn,Bhimji:2026OmniLearned}. FMs have also been applied to calorimeter simulation~\cite{CardonaGiraldo:2026} and to cross-domain transfer from collider jets to neutrino interactions~\cite{Krzmanc:2026fdw}, the latter demonstrating that learned representations can survive a change of both experiment and physics process. Most of these studies operate on high-level or partially reconstructed inputs, including jet constituents, particle-flow candidates, and four-vectors, which are themselves produced by conventional reconstruction chains. At a lower reconstruction level, Ref.~\cite{Giroux:2026ReadoutFM} introduces a proto-FM operating directly on the pixel and timing readout of an imaging Cherenkov detector, fine-tuned for particle identification and noise filtering, although its inputs are hit sequences already associated with a single track.

Substantial effort has also been devoted to machine learning (ML)-based charged-particle tracking, largely motivated by the TrackML challenges~\cite{Amrouche:2019wmx,Amrouche:2021nbs}. Graph neural networks, in which detector hits are treated as nodes and candidate track segments as edges, have been developed for this task, including Exa.TrkX~\cite{Ju:2021} and EggNet~\cite{Calafiura:2024}. Transformer-based~\cite{Miao:2024} and state-space-model-based~\cite{Jiang:2025} tracking approaches have also been explored. Related ML methods, including MLPF~\cite{Pata:2021MLPF} and HGPflow~\cite{Kakati:2025HGPflow}, perform particle-flow reconstruction by combining reconstructed tracks with calorimeter information. Although these models can achieve strong performance, most are optimized for a particular reconstruction objective using task-specific labeled samples.

The Foundation Model for Nuclear and Particle Physics (FM4NPP)~\cite{Park:2025FM4NPP} is a pretrained model designed to learn transferable representations from low-level detector data through self-supervised learning and to support multiple downstream reconstruction tasks. It is pretrained on spacepoints from the sPHENIX Time Projection Chamber (TPC)~\cite{Li:2026TPCpp10M}. In this work, we evaluate FM4NPP on several tracking-related downstream tasks and compare its performance with conventional reconstruction algorithms.

\section{Methods}

The TPC of the sPHENIX experiment has 48 concentric cylindrical readout layers covering a radius of 30--78 cm. It is the primary tracking detector of sPHENIX and covers the full azimuth around the nominal interaction region with a pseudorapidity range of $|\eta|<1.1$. Events were generated with approximately 10 million proton-proton collisions at a center of mass energy of $\sqrt{s}=200$ GeV using \pythia~\cite{Bierlich:2022pfr} and simulated with the full sPHENIX geometry through \geant~\cite{AGOSTINELLI2003250}. 
The simulation accounts for the gas ionization, electron drift in the TPC, and digitization and includes a realistic pad plane readout implementation. Secondary interactions, multiple scattering, and energy loss are handled in \geant. Clusters of TPC hits, referred to as spacepoints, are used as the input for FM training. The input dataset is described in detail in Ref.~\cite{LI2026112393}.

The FM4NPP architecture is based on the Mamba sequence model~\cite{gu2024mambalineartimesequencemodeling}. For pretraining, the TPC is voxelized in $(\eta,\phi,r)$ space, and the spacepoints are hierarchically serialized into sequences. The model is then trained in a self-supervised manner using unlabeled spacepoint data, with a $k$-Next-Nearest-Neighbor prediction task as the pretraining objective. In this work, we compare results derived from the \texttt{m6} model which was trained on nearly 200M parameters. Supervised downstream adapters use the learned spacepoint representations to perform track finding and particle identification while the pretrained FM4NPP backbone remains frozen. The downstream adapter architectures for the labeled training are chosen uniquely for each task; the track finding architecture utilizes transformer decoder layers responsible for aggregating spacepoint embeddings and two separate multilayer perceptron (MLP) heads to produce a classification score, while the PID architecture uses a transformer decoder and single MLP classifier.

Truth particles are matched to reconstructed tracks using spacepoint identifiers (IDs). Each truth-level TPC spacepoint is assigned a unique ID, and each reconstructed spacepoint retains the ID of the truth-level spacepoint from which it originates. Matches between truth particles and reconstructed tracks are identified using the double-majority method, which is commonly used in track-reconstruction studies such as the TrackML challenge~\cite{Amrouche:2019wmx,Amrouche:2021nbs}. The double majority method requires that at least 50\% of the reconstructed track spacepoints originate from a given truth particle, and that at least 50\% of the truth particle's spacepoints are included in the reconstructed track. This ensures a one-to-one correspondence between the reconstructed track and truth particle, and penalizes the reconstruction algorithm for finding only fragments of the truth particle.

The track finding performance of the FM-based reconstruction is compared to that of the Cellular Automaton-based (CA-based) seeder~\cite{Osborn:2021zlr}. The CA algorithm finds track seeds by creating a directed graph using the TPC spacepoints where the outgoing edges are assigned to spatially-close neighbors in adjacent layers, creating triplets. The graph is pruned to keep only pairs of clusters that are connected in both directions, producing chains of clusters that form track seeds. These seeds are typically composed of less than 16 layers of the TPC to reduce the CPU combinatorial burden that arises when extending many long track seeds. The track seeds are subsequently extended using a Kalman Filter based track propagation algorithm~\cite{Fruhwirth:1987fm} to add additional TPC spacepoints to the original track seeds found by the CA algorithm. Tracks reconstructed by both the CA-based and FM-based methods are matched to truth particles using the same double majority criterion described above.

The particle identification (PID) performance of the FM is compared to a traditional ionization-energy-loss (\dedx) method. This method uses Gaussian Process Regression (GPR) to fit the mean of the \dedx distribution as a function of track momentum for each species. Both methods are cast as a
one-vs-rest classification among the three dominant charged hadron species, pions, kaons, and protons ($\pi/K/p$). Their performance is compared using receiver operating characteristic (ROC) curves in the momentum interval $p \in [0.8,1.2)$~GeV/$c$, where the \dedx{} responses of the three species overlap strongly and PID separation in the sPHENIX TPC is expected to be most challenging because the precision of the measured ionization energy loss is limited by the available track length within the physical radial extent of the sPHENIX TPC and by the ionization properties of the TPC gas.

To compare the GPR and FM methods, several conventions are fixed. Both methods are evaluated on the same set of TPC tracks reconstructed by the conventional algorithm, each matched to the truth particle that contributes the most spacepoints; only tracks with at least 20 spacepoints that are matched to primary particles are used. Tracks are grouped charge-inclusively based on their PID and in bins of the track momentum. The classification population is restricted to the three $\pi/K/p$ hypotheses, so for a given species the negative (``rest'') sample is the other two species. The per-track score that is thresholded to trace the ROC curve is constructed identically for both methods and normalized over the same three hypotheses. For the FM, the model emits a per-point class probability that is averaged over the points of the reconstructed track to give a track-level probability $\mathrm{prob}_c$ for each class $c$, which is then renormalized over the three classes, $\mathrm{score}_c = \mathrm{prob}_c / \sum_{k\in\{\pi,K,p\}} \mathrm{prob}_k$. For the GPR method, a per-species likelihood $L_c = 1/\lvert dE/dx - \mu_c(p)\rvert$ is formed from the deviation of the measured \dedx\ from the fitted band mean $\mu_c(p)$, multiplied by a per-momentum-bin class prior $\pi_c(p)$ equal to the species fraction in that momentum bin, and normalized over the three hypotheses $\mathrm{score}_c = L_c\,\pi_c(p) / \sum_{k} L_k\,\pi_k(p)$. A single ROC routine is applied to both sets of scores so that the only difference between the two curves is the score being thresholded.

\begin{figure}[tbh]
    \centering
    \includegraphics[width=\linewidth]{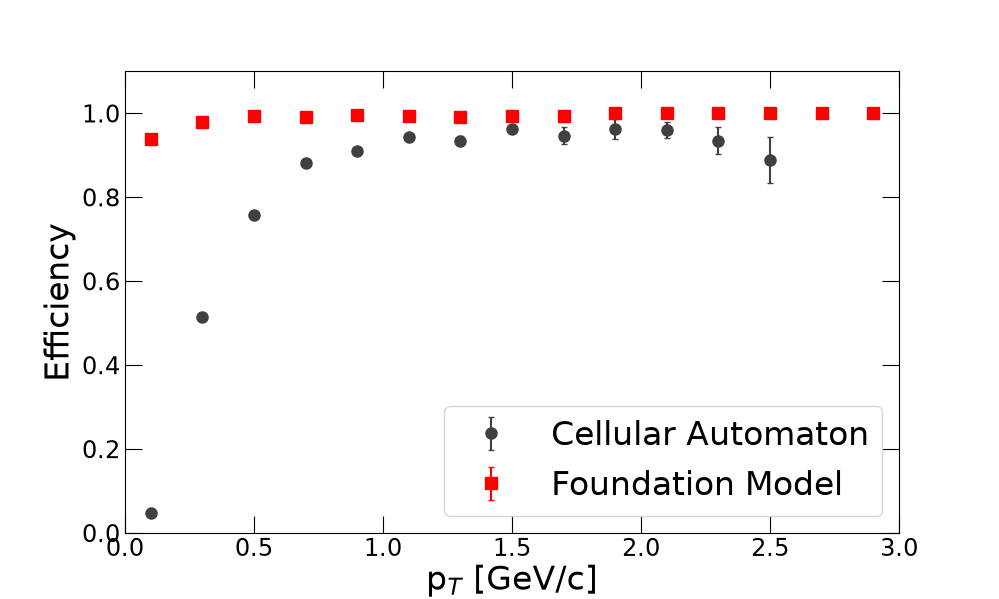}
    \caption{The track finding efficiency for the FM and the Cellular Automaton based seeding is shown as a function of $p_T$. The FM based seeding algorithm surpasses the CA seeder performance at low $p_T$ by a significant margin.}
    \label{fig:efficiency}
\end{figure}

\begin{figure}[tbh]
    \centering
    \includegraphics[width=1\linewidth]{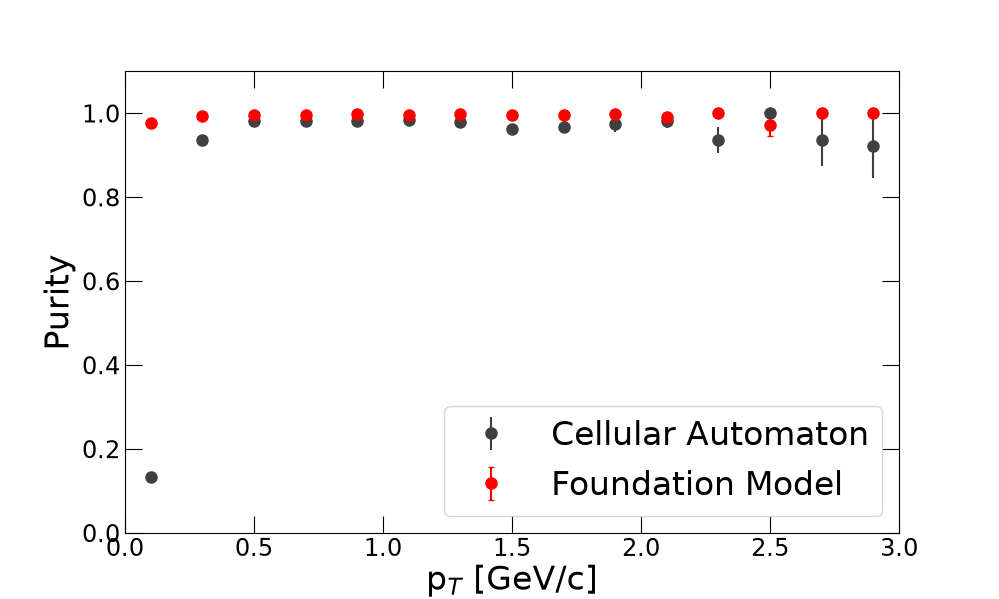}
    \caption{The track finding purity for the FM and the Cellular Automaton based seeding is shown as a function of $p_T$.}
    \label{fig:purity}
\end{figure}

\section{Results}

\begin{figure*}
    \centering
    \includegraphics[width=\linewidth]{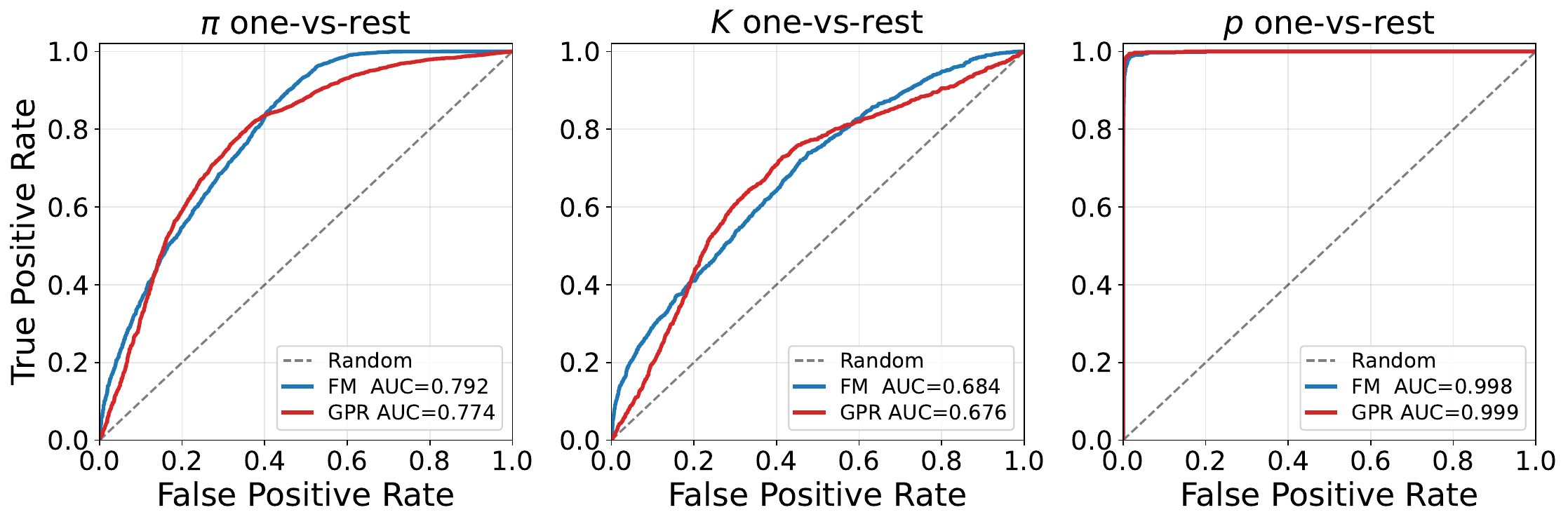}
    \caption{One-vs-rest PID ROC curves for pions (left), kaons (middle), and protons (right) in the momentum slice $p \in [0.8, 1.2)$~GeV/$c$, comparing the FM4NPP model (blue) with the traditional GPR \dedx\ baseline (red). The dashed diagonal denotes random performance, and the area under each curve (AUC) is quoted in the legend. Both methods are evaluated on the same reconstructed tracks matched to primary particles. Tracks are charge-folded and the classification is restricted to the $\pi/K/p$ hypotheses for both methods.}
    \label{fig:pid_roc}
\end{figure*}

The track finding efficiency is shown in Fig.~\ref{fig:efficiency} for both the FM-based reconstruction and the traditional CA-based seeding algorithm. At relatively high $p_T$ both algorithms are quite efficient because the tracks are nearly straight in the magnetic field, making pattern recognition and track following simpler. However, at low $p_T$ ($p_T \lesssim 1.0$) the FM-based reconstruction significantly outperforms the traditional algorithm. Current track finding algorithms typically perform poorly at low $p_T$ due to larger effects of multiple scattering and a higher curvature of the track. Both effects result in pattern recognition becoming more difficult, since there are potentially large deflections of the track and the space points traverse larger regions in azimuth. Additionally, depending on the radial extent of the tracking system and the magnetic field strength, tracks begin to loop and leave nearly identical trajectories at multiple locations in z. Typical reconstruction algorithms are focused on maximizing performance in higher $p_T$ regimes above the threshold for these ``looper'' tracks. Efforts exist to develop standalone algorithms that are responsible for reconstructing loopers~\cite{Battisti:2024nqq}. The benefit of the FM is demonstrated in that a single architecture is capable of reconstructing tracks across the entire $p_T$ spectrum with extremely high efficiency. The training and architecture scheme of the FM are chosen in such a way to learn the patterns of all track momenta, since the same underlying physical principles apply across the full $p_T$ range.

Complementing the efficiency, the purity of the FM-based and CA-based seeding algorithms is shown in Fig.~\ref{fig:purity}. Both algorithms have a very high purity except at low $p_T$, where the CA seeder does not join consecutive loops for particles that loop within the sPHENIX magnetic field with high efficiency. In contrast, the FM has a purity of nearly unity even at low $p_T$ ($p_T \lesssim 0.5$). This shows that the FM finds the correct track and also that it does not duplicate tracks, even for low $p_T$ loopers, which may leave many bending track patterns along the $z$ axis in a single event within the TPC.

The results of the PID separation are shown in Fig.~\ref{fig:pid_roc}. Protons are cleanly separated by both methods with an AUC $\geq 0.998$ for both, reflecting the large \dedx\ separation of protons from the lighter species in this momentum range. For pions the FM slightly outperforms the GPR baseline (AUC 0.792 versus 0.774), while for kaons---the most challenging case, as they sit between the pion and proton bands---the two methods are comparable, with the curves crossing and the FM area under the curve slightly higher (0.684 versus 0.676). The PID performance is primarily determined by the physical characteristics of the TPC; for example, a TPC with a larger radius and ionizing gas that liberates more electrons will have larger \dedx per track, improving the possibility of particle identification separation. The sPHENIX TPC was designed to mitigate effects from space charge distortions and not for PID performance. The FM performance being comparable to, and slightly better than, the traditional GPR performance is a powerful result when considered in the context of the FM pretraining objective focusing on track finding and spatial pattern recognition. Despite this, the FM learned features in the cluster correlations amongst the tracks to discern what the energy loss characteristics in the TPC were and was able to perform \dedx based separation comparably to the GPR method without explicit training.

Several directions remain for future work. The FM studied here was trained using data from a single detector subsystem, whereas nuclear and particle physics experiments contain multiple subsystems that provide complementary information about reconstructed particles. An experiment-wide foundation model could incorporate information from tracking detectors, calorimeters, and other detector systems. For example, incorporating calorimeter information could enable the reconstruction of particle-flow-like objects that combine tracking and deposited-energy measurements. The broader FM4NPP roadmap identifies the development of AI-ready data infrastructure, scaling to larger and multimodal models, knowledge transfer across experiments and facilities, and the integration of scientific reasoning and tool-use capabilities as important future directions~\cite{Ren:2026fm4npproadmap}. Transfer learning across different physics domains has already shown promising results~\cite{Krzmanc:2026fdw}. Further studies using spacepoint data from other detectors, collision systems, and experimental conditions are needed to determine the generalizability of the learned representations. Furthermore, selecting pretraining objectives that are more generalizable, such as one that may include energy loss for particle identification, may improve the performance of the FM for a wider range of downstream tasks.

\section{Conclusions}
We have evaluated an FM-based approach to low-level particle reconstruction using simulated sPHENIX TPC data. The FM-based track-finding method substantially improves the reconstruction efficiency relative to the conventional Cellular Automaton-based method at low $p_T$ while maintaining high purity across the studied momentum range. The improvement is particularly pronounced for low-$p_T$ looping tracks, which are challenging for conventional reconstruction algorithms. At higher $p_T$, both methods achieve similarly high tracking efficiency and purity.

The same frozen pretrained backbone can also support particle identification through a supervised downstream adapter. The FM-based particle identification method achieves performance similar to that of the conventional GPR-based \dedx{} method for kaons and protons in the selected momentum interval, and slightly better pion discrimination. Notably, the pretraining objective did not explicitly involve energy or ionization-energy-loss prediction. These results indicate that a shared pretrained backbone could support several reconstruction tasks using appropriate downstream adapters, potentially simplifying data analysis and extending the reach of measurements, particularly at low $p_T$.

\begin{acknowledgments}
The authors would like to express their sincere gratitude to the sPHENIX collaboration for sharing the simulation data and experimental knowledge, as well as Hong Ma and James Dunlop from Brookhaven National Laboratory for their valuable support and feedback. Additionally, the authors thank Chris Pinkenburg, Michael Peters, and Tony Frawley for their expertise and feedback on the sPHENIX computing and tracking software. This work was partially supported by the Laboratory Directed Research and Development (LDRD) Program at Brookhaven
National Laboratory, LDRD 25-045 and 27-025, which is operated and
managed for the U.S. Department of Energy (DOE) Office of
Science by Brookhaven Science Associates under contract
No. DE-SC0012704. Shuhang Li was partially supported
by the DOE Office of Science through the Office of Nu-
clear Physics under Award No. DE-FG02-86ER4028
\end{acknowledgments}

\bibliography{bibliography.bib}

\end{document}